# The Poincaré conjecture for digital spaces. Properties of digital n-dimensional disks and spheres.

Alexander V. Evako

Volokolamskoe Sh. 1, kv. 157, 125080, Moscow, Russia
Tel/Fax: 495 158 2939, e- mail: evakoa@mail.ru.

Abstract.
Motivated by the Poincaré conjecture, we study properties of digital n-dimensional spheres and disks, which are digital models of their continuous counterparts. We introduce homeomorphic transformations of digital manifolds, which retain the connectedness, the dimension, the Euler characteristics and the homology groups of manifolds. We find conditions where an n-dimensional digital manifold is the n-dimensional digital sphere and discuss the link between continuous closed n-manifolds and their digital models.

1. Introduction

A digital approach to geometry and topology plays an important role in analyzing n-dimensional digitized images arising in computer graphics as well as in many areas of science including neuroscience, medical imaging, industrial inspection, geoscience and fluid dynamics. Concepts and results of the digital approach are used to specify and justify some important low-level image processing algorithms, including algorithms for thinning, boundary extraction, object counting, and contour filling [1-9,11,16-17].
It seems desirable to study properties of digital n-manifolds in a fashion that more closely parallels the classical approach of algebraic topology in order to find out, how far the fundamental distinction between continuous and digital spaces due to different cardinality restricts a direct modification of continuous tools to digital models on one hand and how effectively the digital approach can be applied to solve classical topology problems on the other hand. As an example, we consider the Poincaré conjecture about the characterization of the three-dimensional sphere amongst 3-manifolds.
The material to be presented below begins with the description of computer experiments that provide a reasonable background for introducing digital spaces as simple graphs with a topological structure (section 2). Then we remind basic definitions of digital normal n-dimensional spaces (n-spaces) (section 3).
In sections 4, we introduce and study properties of digital n-disks and n-spheres, which are similar to properties of their continuous counterparts. We introduce homeomorphic transformations of digital n-manifolds, which retain their basic topological features. It is proven that a digital n-sphere converts into the minimal one by homeomorphic transformations and that a digital n-sphere without a point is homotopic to a point. In sections 5, we study properties of compressed digital n-manifolds. Finally, (sections 6 and 7) we consider conditions including Poincaré conjecture about the characterization of the digital two-, three and n-spheres amongst digital two-, three and n-manifolds.

2. Computer experiments as the basis for digital spaces.

An important feature of this approach to the structure of digital spaces is that it is based on computer experiments which results can be applied to computer graphics and animations.

The following surprising fact is observed in computer experiments modeling deformation of continuous surfaces and objects in three-dimensional Euclidean space [13]. Suppose that $S_1$ is a surface (in general, a one- two- or thee-dimensional object) in Euclidean space $E^3$. Tessellate $E^3$ into a set of unit cubes, pick out the family $M_1$ of unit cubes intersecting $S_1$ and construct the digital space $D_1$ corresponding to $M_1$ as the intersection graph of $M_1$ [10]. Then reduce the size of the cube edge from 1 to 1/2 and using the same procedure, construct $D_2$. Repeating this operation several times, we obtain a sequence of digital spaces $D=\{D_1,D_2,.. D_n,...\}$ for $S_1$. It is revealed that the number p exists such that for all m and n, m>p, n>p, $D_m$ and $D_n$ can be turned from one to the other by four kinds of transformations called contractible. Moreover, if we have two objects $S_1$ and $S_2$, which are topologically equivalent, then their digital spaces can be converted from one to the other by contractible transformations if a division is small enough. It is reasonable to assume that digital spaces contain topological and perhaps geometric characteristics of continuous surfaces and contractible transformations digitally model mapping of continuous spaces.

3. Preliminaries.

By a digital space G we mean a simple graph G=(V,W) with a finite or countable set of points $V=(v_1,v_2,...v_n,...)$, together with a set of edges $W = ((v_pv_q),....)\subseteq V\times V$ provided that $(v_pv_q)=(v_qv_p)$ and $(v_pv_p)\notin W$ [7]. Points $v_p$ and $v_q$ are called adjacent if $(v_pv_q)\in W$. We use the notations $v_p\in G$ and $(v_pv_q)\in G$, if $v_p\in V$ and $(v_pv_q)\in W$ respectively if no confusion can result. A graph K is called complete if every two points of K are adjacent. A graph $v\oplus G$ is a graph where point v is adjacent to all points of graph G. A graph $v\oplus G$ is called the cone of graph G. $H=(V_1,W_1)$ is a subspace of G=(V,W) if $V_1\subseteq V$ and $W_1=W\cap(H\times H)$ It means that points $v_p,v_q\in H$ are adjacent in H if and only if they are adjacent in G. In other words H is an induced subgraph of the graph G. Let G be a digital space and v be a point of G. The subspace U(v) containing v as well as all its neighbors is called the ball of point v in G. The subspace O(v)=U(v)-v containing only neighbors of v is called the rim of point v in G.
The subspace $O(v_1,v_2,v_3,....v_P,)$ formed by the intersection of $O(v_1)$, $O(v_2)$, $O(v_3)$, ... $O(v_P)$ is called the joint rim of points $v_1, v_2, v_3,.... v_P$.
$O(v_1,v_2,v_3,....v_P)=O(v_1)\cap O(v_2)\cap O(v_3)\cap...\cap O(v_P)$.
Digital spaces can be transformed from one into another in variety of ways. Contractible transformations of digital spaces [12,13] seem to play the same role in digital topology as homotopy in algebraic topology. It can be checked directly that if two standard two- or three-dimensional spaces are homotopically equivalent in algebraic topology, then a digital counterpart of one of them can converted into a digital counterpart of the other by contractible transformations. A space G is called contractible if can be converted to a single point by a sequence of contractible transformations.
If a space G can be obtained from a space H by a sequence of contractible transformations [12], then we say that G is homotopic to H, G~H. Homotopy is an equivalence relation among digital spaces. Contractible transformations retain the Euler characteristic and homology groups a graph.
 Subgraphs A and B of a graph G are called separated or non-adjacent if any point in A is not adjacent to any point in B.
The join $G\oplus H$ of two spaces G=(X,U) and H=(Y,W) is the space that contains G, H and edges joining every point in G with every point in H [6,7]. In graph theory, this operation is also called the join of two graphs [10]. Remind the isomorphism of digital spaces. Note that the isomorphism of digital spaces is the isomorphism of graphs [10] if we see a digital space as a graph. A complete space K is a complete graph, a set of s points every two of which are adjacent to each other. A digital space G with a set of points $V=(v_1,v_2,...v_n)$ and a set of edges $W=((v_p v_q),....)$ is said to be isomorphic to a digital space H with a set of points $X=(x_1,x_2,...x_n)$ and a set of edges $Y=((x_p x_q),....)$ if there exists one-one onto correspondence f: $V \to X$ such that $(v_i v_k)$ is an edge in G iff $(f(v_i)f(v_k))$ is an edge in H. Map f is called an isomorphism of G to H.

We write G=H to denote the fact that there is an isomorphism of G to H.

Let G and H be digital spaces and A and B be their subspaces, $A \subseteq G$, $B \subseteq H$. If A and B are isomorphic, then the space G#H obtained by identifying points in A with corresponding points in B is said to be the connected sum of G and H over A (or B).

Definition 3.1.

A normal digital 0-dimensional space is a disconnected graph $S^0(a,b)$ with just two points a and b. For n>0, a normal digital n-dimensional space is a nonempty connected graph $G^n$ such that for each point v of $G^n$, O(v) is a normal finite digital (n-1)-dimensional space [6,7].

By this definition, if G is the empty space, then its dimension is (−1). The normal digital 0-dimensional space is called the normal digital 0-dimensional sphere.

Proposition 3.1 [6,7].

Let $G^n$ and $H^m$ be normal n- and m-dimensional spaces. Then their join $G^n \oplus H^m$ is a normal (n+m+1)-dimensional space.

Proposition 3.2.

Let $G^n$ be a normal n-dimensional space and $K(v_1,v_2,...v_k)$ $1 \le k \le n$ is a complete subspace of $G^n$. (every two points are adjacent).

Then the joint rim $O(v_1,v_2,...v_k)$ of these points (the mutual adjacency set) is a normal (n-k)-dimensional space.

For any complete subspace $K(v_1,v_2,...v_k)$ there is a the complete subspace $K(v_1,v_2,...v_k,v_{k+1},...v_n,v_{n+1})$ [7].

Proposition 3.3.

If $G^n$ and $H^n$ are normal n-dimensional spaces, and $H^n$ is a subspace of $G^n$, then $H^n=G^n$.

Proposition 3.4.

- The cone $v \oplus G$ of any space G is a contractible space.
- Let G and v be a contractible graph and any its point. Then graphs O(v) and G-v are homotopic and G-v converts into O(v) by contractible deleting of points in any suitable order [12].

Remind that in classical approach of algebraic topology, loosely speaking, G is an n-dimensional manifold without boundary if each point has a neighborhood homeomorphic to the open n-ball in the n-dimensional Euclidean space. Normal n-dimensional digital spaces were designed to model continuous n-manifolds. In this paper we deal only with manifolds. Since we only use digital normal n-dimensional spaces we will use the word n-space for a digital normal n-dimensional space.

4. Properties of n-spheres and n-disks.

A zero-ball is a single point v, a zero-sphere $S^0(a,b)$ is a disconnected graph with just two points a and b. A one-sphere $S^1$ is a connected graph such that for each point v of $S^1$, its rim O(v) is a zero-sphere $S^0$ (fig. 1). A one-disk $D^1$ is a connected graph $S^1$-v obtained from a one-sphere $S^1$ by the deleting of a point v.

Formally, a zero-dimensional disk is a single point. Remind that subgraphs A and B of a graph G are called separated or non-adjacent if any point in A is not adjacent to any point in B.

Lemma 4.1.

The minimal one-sphere consists of four points, $S^1_{min} = S^0(a,b) \oplus S^0(c,d)$.

The minimal one-disk consists of three points, $D^1_{min} = a \oplus S^0(b,c)$

Any zero-sphere $S^0$ belonging to a one-sphere $S^1$ divides $S^1$ into two separated parts. If $S^1$ is a one-sphere and $S^0$ is a zero-sphere belonging to $S^1$, then $S^1$ is the union $A \cup S^0 \cup B$ where subspaces A and B are separated and subspaces $A \cup S^0$ and $B \cup S^0$ are one-disks.

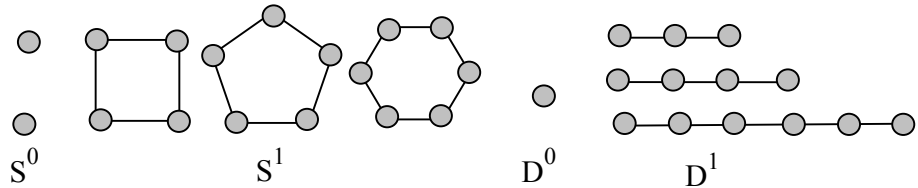

$S^0$      $S^1$      $D^0$      $D^1$

Figure 1. Zero- and one-dimensional spheres $S^0$ and $S^1$ and zero- and one-dimensional disks $D^0$ and $D^1$.

The properties are checked directly.

To define n-disks and n-spheres, we will use a recursive definition. Suppose that we have defined k-disks and k-spheres for dimensions $1 \leq k \leq n-1$.

Definition 4.1.

( a ) A connected space N is called a closed n-manifold if the rim of any point of N is an (n-1)-sphere. [6,7].

( b ) A connected space N is called an n-manifold with boundary ∂N, if there exists a closed n-manifold M and a point v in M such that M-v is isomorphic to N. The subspace O(v) is called the boundary of N, the subspace IntN=N-∂N is called the interior of N [8]. For a closed n-manifold we omit the word closed if no confusion can result.

Obviously, ∂N is an (n-1)-sphere.

Definition 4.2.

( a ) An n-manifold D with boundary is called an n-disk if D is a contractible space (that is D can be converted to a point by contractible transformations) (fig. 2).

( b ) Let D and C be n-disks such that ∂D and ∂C are isomorphic, ∂D=∂C. The space D#C obtained by identifying each point in ∂D with its counterpart in ∂C is called an n-sphere. Obviously, S is the connected sum of D and C over ∂D.

As it follows from definitions 4.1 and 4.2, the join $v \oplus S$ of a point v and an (n-1)-sphere S is an n-disk. For example, the ball $U(v)=v \oplus O(v)$ of any point v in a closed n-manifold N is an (n-1)-disk.

Definition 4.3.

( a ) Let N be an n-manifold and D be an n-disk belonging to N. We say that n-disk D is replaced with a point v (or with an n-disk $v \oplus \partial V$) if the space M=N+v-IntD is obtained by joining point v with any point in ∂D and deleting from N points belonging to IntD.

( b ) Let N be an n-manifold and v be a point belonging to N. Let D be an n-disk such that O(v) is isomorphic to ∂D (D does not belong to N). We say that point v (or the n-ball U(v)) is replaced with n-disk D if the space M=N-v+IntD is obtained by identifying any point in ∂D with its counterpart in O(v) and deleting point v.

We call replacings (a) and (b) homeomorphic transformations or h-transformations.

In fact, both operations are the replacings of n-disks by n-disks. The replacing of n-disks in an n-manifold N is an application of contractible transformations [12] of digital spaces to n-manifolds. It follows from definition 4.3 that h-transformations are represented by a sequence of contractible transformations of digital spaces that retain such properties of digital spaces as the Euler characteristic and the homology groups [13,14].

Definition 4.4.
( a ) The join $S^n_{min}=S^0_1\oplus S^0_2\oplus\ldots S^0_{n+1}$ of (n+1) copies of the zero-sphere $S^0$ is called the minimal n-sphere.
( b ) The join of a point v and the minimal (n-1)-sphere $S_{min}$ is called the minimal n-disk, $U_{min}=v\oplus S_{min}$ (fig. 3)

Lemma 4.2.
( a ) The join $S^0(u,v)\oplus D$ of a zero-sphere $S^0(v,u)$ and an n-disk D is an (n+1)-disk.
( c ) The join $S^0(u,v)\oplus S$ of a zero-sphere $S^0(v,u)$ and an n-sphere S is an (n+1)-sphere.
It follows directly from definitions 4.1 and 4.2

Lemma 4.3.
( a ) An n-sphere S is homeomorphic to the minimal n-sphere $S_{min}$.
Proof.
Suppose that S=D#C. Replace n-disks D and C in S by n-disks $E=v\oplus\partial D$ and $F=u\oplus\partial C$. Then S converts into $S^0(v,u)\oplus\partial D$ where $\partial D$ is an (n-1)-sphere. For the same reason as above, $\partial D\approx S^0(a,b)\oplus\partial E$ where $\partial E$ is an (n-2)-sphere.
Hence, $S\approx S^0(v,u)\oplus S^0(a,b)\oplus\partial E$. Repeat the above transformations until we obtain the minimal n-sphere. □

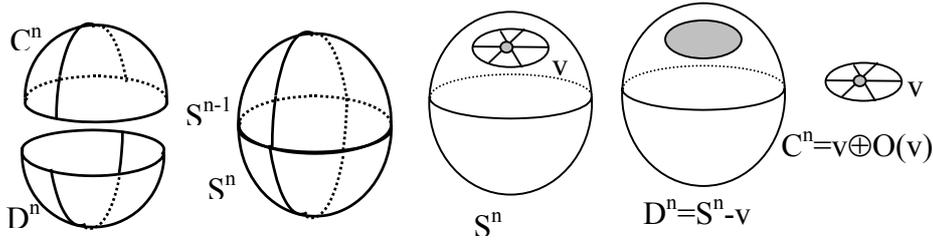

Figure 2. An n-sphere $S^n$ is the connected sum of n-disks $C^n$ and $D^n$. If $S^n$ is an n-sphere, then $D^n=S^n-v$ is an n-disk

Lemma 4.4.
( a ) Let S be an n-sphere and v be a point of S. Then a space S-v obtained from S by the deleting of point v is an n-disk (fig. 4).
Proof.
Note that for $S=S^n_{min}$, the lemma is obvious. Assume that the lemma is valid for S with a number of points |S|=r≤s. Let r=s+1. By definition 4.2, S is the connected sum D#C of n-disks D and C over $\partial D$. With no loss of generality, suppose that a point v belongs to the interior of D, v∈IntD, and |IntC|>1. Replace C by an n-disk $E=x\oplus\partial C$. Then S converts into $S_1=E\#D$ where $|S_1|<s+1$. $S_1$ is the n-sphere by definition 4.2. Therefore, $S_1-v=F$ is an n-disk by the assumption i.e., F is a contractible space. Obviously, x⊆F. Replace x by C. Then F converts into S-v. This is an n-manifold with boundary. Since this replacing is a sequence of contractible transformations, then S-v is contractible. Therefore, S-v is an n-disk by definition 4.2. □

Lemma 4.5.
Let S be an n-sphere and D be an n-disk belonging to S. Then S-IntD is an n-disk.
Proof.
We have to prove that S-IntD is a contractible space and an n-manifold with boundary. Note that if D is the ball U(v) of a point v, then IntD=v and C=S-IntD=S-v is an n-disk by lemma 4.4.
( a ) Let us prove that S-IntD is a contractible space. Suppose that IntD contains more than one point. Take an n-disk $E=v\oplus\partial D$ where a point v does not belong to S (fig. 5). Then a space D#E over ∂D is an n-sphere by definition 4.2. Let a point u belong to IntD. Then F=D#E-u is an n-disk by lemma 4.4. Therefore, F is a contractable space. By proposition 4.4, F-v converts into O(v)=∂D by the contractible deleting of points. Therefore S-v converts into S-IntD by the same contractible deleting of points. Hence C=S-IntD is a contractible space.

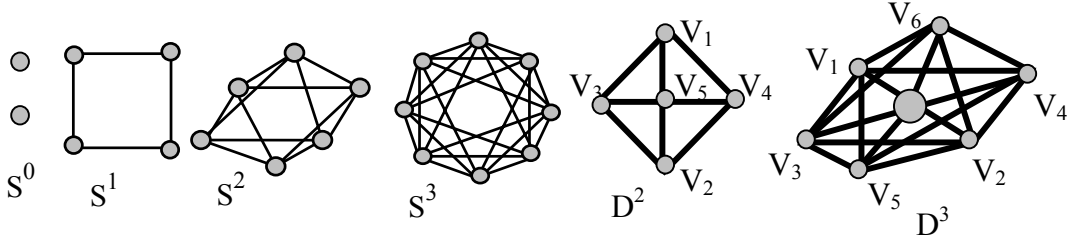

Figure 3. Minimal spheres and disks.

( b ) To prove that S-IntD is an n-manifold with boundary let us use the induction. For n=1, the lemma is plainly true. Assume as an induction hypothesis that the lemma is valid whenever n<s+1. Let n=s+1. Consider a point x belonging to ∂D. The rim A(x) of x in S is an (n-1)-sphere. Obviously A(x)=(S∩D)∪(S∩C). The rim B(x)=(S∩D)+v of x in D#E is also an (n-1)-sphere containing point v. The rim of x in D is F(x)=B(x)-v=S∩D. According to lemma 4.4, F(x) is an (n-1)-disk. The rim G(x) of x in C is A(x)-IntF(x). Therefore, G(x)=A(x)-IntF(x) is an (n-1)-disk by the induction hypothesis. The rim any point y belonging to C and not belonging to ∂D is the same as the rim of y in S i.e., an (n-1)-sphere. Therefore, C is an (n-1)-manifold with boundary. Hence, C is an (n-1)-disk by definition 4.2. □

The following lemma was proven in [8] for normal n-dimensional manifolds with boundary which are not necessarily n-manifolds. Here we prove the same result for an n-manifold.

Lemma 4.6.

Let N be an n-manifold with boundary ∂N and v be a point belonging to ∂N. Then O(v) is an (n-1)-disk.

Proof.

By definition 4.2, there is a closed n-manifold M containing a point x such that M-x=N. Then O(x)=∂N. Let a point v belong ∂N, B(v) be the rim of point v in M and O(v) be the rim of point v in N. Then O(v)=B(v)-x. Since B(v) is an (n-1)-sphere, then O(v) is an (n-1)-disk by lemma 4.4. □

Lemma 4.7.

h-Transformations convert an n-manifold into an n-manifold.

Proof.

With no loss of generality, consider a closed n-manifold N. Let M=gN be a space obtained from N by h-transformation g. We have to prove that for any point v of M, the rim O(v) is an (n-1)-sphere. Let D be an n-disk belonging to N. Suppose that M=N+v-IntD is obtained by connecting a point v with any point in ∂D and deleting from N all points belonging to IntD. We have to prove that the rim of any point of M is an (n-1)-sphere. Let a point x belong to ∂D, B(x) be the rim of x in M, O(x) be the rim of x in N and A(x)=O(x)∩D be the rim of point x in D. Since x is a boundary point of D, then A(x) is an (n-1)-disk according to lemma 4.6. Since O(x) is an (n-1)-sphere and A(x)⊆O(x), then C(x)=O(x)-IntA(x) is an (n-1)-disk by lemma 4.5. Obviously, ∂C(x)=∂A(x). Then B(x) is the connected sum of the n-disk C(x) and the n-disk E=v⊕∂C(x) over ∂C(x), where E=v⊕∂C(x). Hence, B(x)=C(x)#E is the (n-1)-sphere by definition 4.2.

If a point y does not belong to D, then its rim does not change and remains an (n-1)-sphere. The rim of point v is an (n-1)-sphere ∂D. Therefore, M is a closed n-manifold by definition 4.1.

Suppose that the boundary ∂D of an n-disk D is isomorphic to the rim O(v) of some point v of N. In the same way as above, we can prove that M=N-v+IntD obtained by identifying any point in ∂D with its counterpart in O(v) and deleting point v is an n-manifold. □

h-Transformations can be applied to an n-disk C if for an n-disk D in definition 4.3(a), IntD⊆IntC or a point v∈IntC in definition 4.3(b).

Lemma 4.8.

h-Transformations convert an n-disk into an n-disk.

Proof.

Let C=gD be a space obtained from an n-disk D by an h-transformation g. We have to prove that C is a contractible space and an n-manifold with boundary.

Since D is a contractible space and g is a sequence of contractible transformations, then C is contractible.

To prove that C is an n-manifold with boundary, note ∂C=∂D. Let M=D+v be the space obtained from D by the gluing of a point v to D in such a manner that O(v)=∂D. Obviously, M is a closed n-manifold and g is an h-transformation of M. According to lemma 4.7, gM is a closed n-manifold. Therefore, C=gM-v=gD is an n-manifold with boundary. Hence, C is an n-disk. □

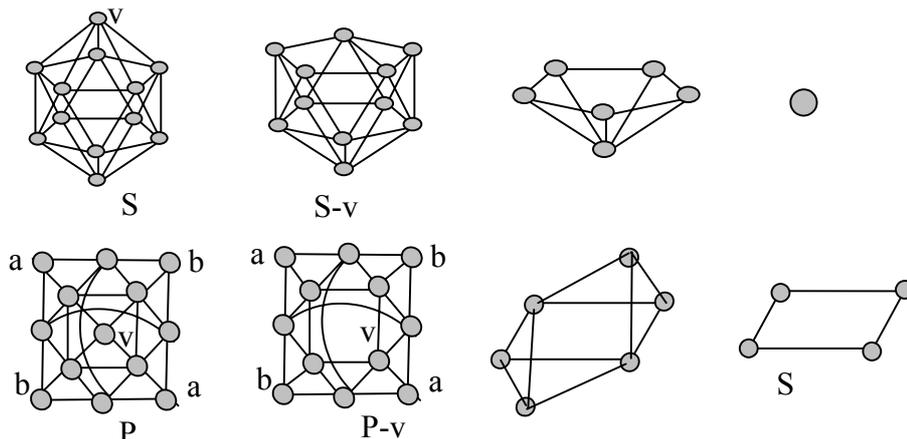

Figure 4. S is a two-sphere, S-v is a two-disk which is homotopic to a point. P is a two-dimensional projective plane, P-v is homotopic to a one-sphere S.

Lemma 4.9.

h-Transformations convert an n-sphere into an n-sphere.

Proof.

Let N=gS be the space obtained from an n-sphere S by an h-transformation g. We have to prove that N can be represented as the connected sum of n-disks A and B over ∂A.

Suppose that S is an n-sphere and v is a point belonging to S. Let D be an n-disk such that ∂D is isomorphic to O(v). Suppose that N=gS=S-v+IntD is the space obtained by identifying any point in ∂D with its counterpart in O(v) and deleting point v. Clearly, N is the connected sum of A=S-v and D over ∂D. A is an n-disk by lemma 4.4. Therefore, N=A#D is an n-sphere.

Suppose that S is an n-sphere and D be an n-disk belonging to S. Let the space N=S+v-IntD is obtained by joining point v with any point in ∂D and deleting from S points belonging to IntD. Then N is the connected sum Of A=S-IntD and B=v⊕∂D. A is an n-disk by lemma 4.5, B is an n-disk by definition 2.2. Therefore, N=A#B is an n-sphere by definition 4.2. □

Definition 4.5.
Closed n-manifolds M and N are called homeomorphic if one of them can be obtained from the other by a sequence of h-transformations.
The following corollaries are an easy consequence of previous results.
Corollary 4.1.
A closed n-manifold M is an n-sphere if and only if there is a point v such that M-v is a contractible space.

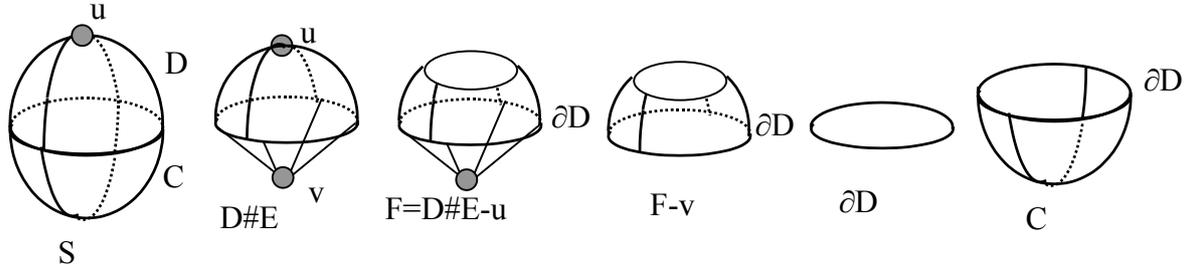

Figure 5. C=S-IntD is a contractible space.

Corollary 4.2.
If a space N is obtained from the minimal n-sphere $S^n_{min}$ by a sequence of h-transformations, then N is the n-sphere.
Figure 4 shows a two-sphere S and a two-dimensional projective plane P. S-v is a two-disk, which is homotopic to a point. P-v is a Mebius band, which is homotopic to a circle.

Lemma 4.10.
Let N be a closed n-manifold. If for adjacent points u and v, the union of their balls U(u)∪U(v) is not an n-disk, then there is a one-sphere S(4) consisting of four points and containing points u and v.
Proof.
Let points v and u be adjacent (fig. 6). Suppose that the union C=U(v)∪U(u) is not an n-disk. Since O(v) and O(u) are (n-1)-spheres, then subspaces A=O(v)-u and B=O(u)-v are (n-1)-disks by lemma 4.4. Suppose that there is no connection between points belonging to IntA and IntB. Then A#B is an (n-1)-sphere and C is an n-disk by definitions 4.1 and 4.2. This contains a contradiction because C is not an n-disk. Therefore, there are points w∈IntA and a∈IntB, which are adjacent. Hence, {v,u,a,w} is the one-sphere. □

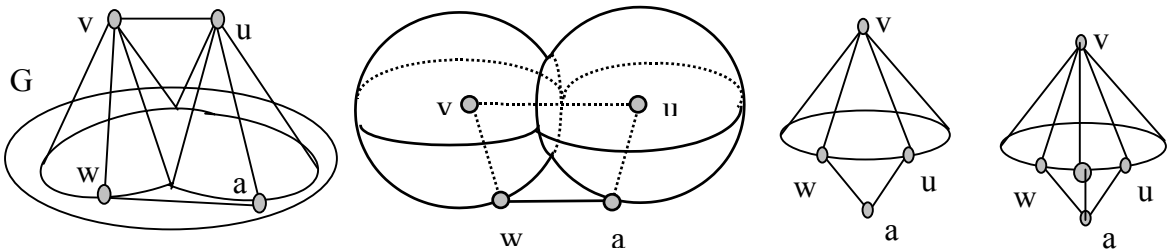

Figure 6. For points v and u there is a one sphere $S^1$ containing these points.

Lemma 4.11.
Let N and M be n-manifolds with the boundary such that N⊆M and ∂N⊆∂M. Then N=M.
The assertion follows from definitions 4.1, 4.2 and proposition 3.3.

5. Two-manifolds.

Obviously, a two-space is a two-manifold.
Theorem 5.1.
Let $H^2$ be a closed two-manifold. If for any one-sphere $S^1$ belonging to $H^2$ there is a two-disk $D^2$ belonging to $H^2$ such that $S^1=\partial D^2$, then $H^2$ is a two-sphere.
Proof.
Replace any two-disk $D^2$ with $v\oplus\partial D^2$. Since this replacing is an h-transformation, then the obtained closed two-manifold $G^2$ is homeomorphic to $H^2$. In $G^2$, any disk $D^2$ is the rim $O(v)$ of some point v. Take adjacent points v and $a_1$. Since the subspace $U(v)\cup U(a_1)$ is not a two-disk, then by lemma 4.10, there is a one-sphere $S^1=\{v,a_1,u,a_3\}$ consisting of four points and containing points v and $a_1$ (fig. 7). Since for $S^1$, there is a two-disk $D^2$ such that $S^1=\partial D^2$ and any $D^2$ is the ball of some point, then there is a point $a_2$ adjacent to all points in $S^1$ i.e., $O(a_2)=S^1$. Consider points v and $a_2$. Since $U(v)\cup U(a_2)$ is not a two-disk, then by lemma 4.10, there is a one-sphere $S^1_1$ consisting of four points and containing points v and $a_2$. Clearly, $S^1_1$ must contain point u i.e, $S^1_1=\{v,a_2,u,a_4\}$  For $S^1_1$, there is a two-disk $D^2$ such that $S^1_1=\partial D^2$. Since any $D^2$ is the ball of some point, then there is a point x adjacent to all points in $S^1_1$. Obviously, either $x=a_3$ or $x=a_1$. Suppose $x=a_3$. Using the above arguments, we find that point u is adjacent to all points of $O(v)$. Therefore, the subspace $U(v)\cup U(u)$ is the two-sphere $S^2=S^0(v,u)\oplus O(v)$. Since $S^2\subseteq G^2$, then $S^2=G^2$ by proposition 3.3. Therefore, $H^2$ is the two-sphere by corollary 4.2. The proof is complete. □
As we can see from the proof of theorem 6.2, the condition "for any one-sphere $S^1$ belonging to a closed two-manifold  there is a two-disk $D^2$ belonging to this manifold such that $S^1=\partial D^2$" is applied only to $G^2$ which is obtained from $H^2$ by the replacing of all two-disk $D^2$ by $v\oplus\partial D^2$.
Figure 8 shows the two-sphere $S^2$ and the two-dimensional projective plane $P^2$. It can be checked directly that for any one-sphere $S^1$ belonging to $S^2$ there is a two-disk $D^2$ such that $\partial D^2=S^1$. For a one-sphere $S^1=\{1,4,7,10\}$ belonging to $P^2$, there is no two-disk $D^2$ such that $\partial D^2=S^1$. In the next theorem we use another condition, which guaranties that $H^2$ is a two-sphere.

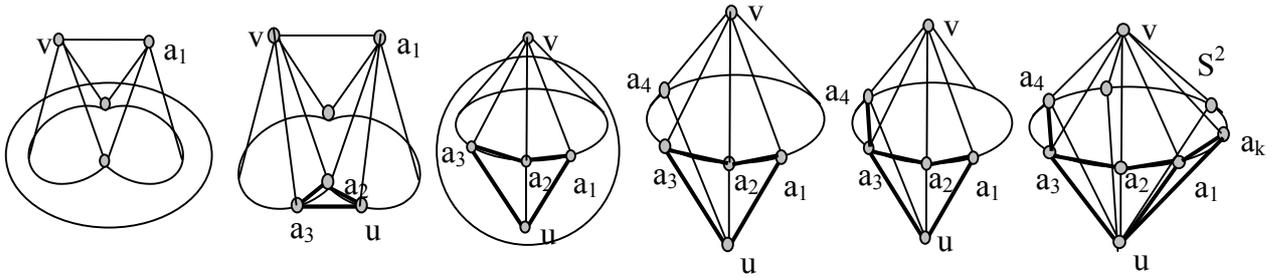

Figure 7. $S^2$ is a subspace of $G^2$. Therefore, $G^2=S^2$.

Definition 5.1.
A one-disk $D^1$ belonging to a two-manifold $H^2$ is called embedded in $H^2$ if there is a two-disk $D^2$ belonging to $H^2$ such that $\partial D^1\subseteq\partial D^2$, $IntD^1\subseteq IntD^2$.
Theorem 5.2.
Let $H^2$ be a closed two-manifold. If any one-disk $D^1\subseteq H^2$ is embedded in $H^2$, then $H^2$ is the two-sphere.
Proof.
Take any two-disk $D^2$ belonging to $H^2$ and replace it with the two-disk $v\oplus\partial D^2$ by the deleting of all points belonging to $IntD^2$ and the connecting of a point v with any point of $\partial D^2$. This is an h-transformation that converts $D^2$ into $U(v)=v\oplus\partial D^2$. Repeat this procedure until any two-disk $D^2$ is the ball of some point v, $D^2=U(v)$. Denote the obtained space by $G^2$. Clearly, $G^2$ is homeomorphic to $H^2$.

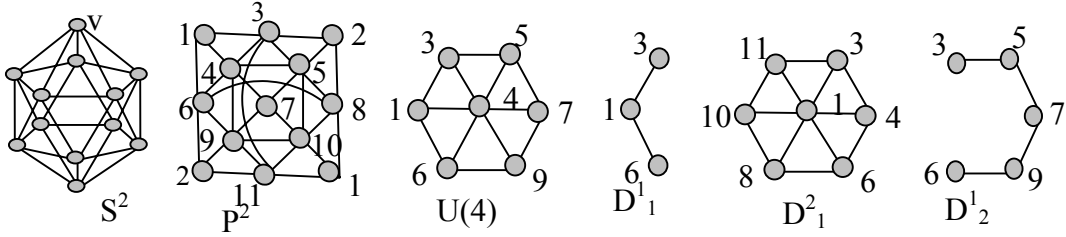

Figure 8. In the two-sphere $S^2$, for any one-sphere $S^1$ there is a two-disk $D^2$ such that $\partial D^2=S^1$.. In the projective plane $P^2$, for a one-sphere $S^1=\{1,4,7,10\}$, there is no two-disk $D^2$ such that $\partial D^2=S^1$.

For a point $v \in G^2$, take some point u belonging to a one-sphere $O(v)$ (fig. 9). Then $S^0(x,y)=O(v,u)=O(v) \cap O(u)$ is a zero-sphere by proposition 3.2, $u \oplus O(v,u)=D^1_1$ is a one-disk belonging to $O(v)$ and $O(v)$-u also is another one-disk $D^1_2$ belonging to $O(v)$ by lemma 4.4. Obviously, $S^0(x,y)=\partial D^1_1=\partial D^1_2$ and $O(v)=D^1_1 \cup D^1_2$. Since $D^1_2$ is embedded, then there is two-disk $D^2_2$ such that $\partial D^1_2 \subseteq \partial D^2_2$ and $IntD^1_2 \subseteq IntD^2_2$. Since $D^2_2$ is the ball of some point $a_1$, $D^2_2=U(a_1)$, then point v necessarily belongs to $\partial D^2_2$ and $O(v,a_1)$ is $S^0(x,y)$. Hence, $O(v)=S^1_{min}=S^0(x,y) \oplus S^0(u,a_1)$. Applying the above arguments to all points in $G^2$, we see that the rim of any point is $S^1_{min}$ consisting of four points. Let the rim of some point v be $S^1_{min}$ consisting of points $\{u,x,a_1,y\}$ and consider the rim of point $a_1$. $O(a_1)$ is also $S^1_{min}$ and contains four points $\{y,v,x,w\}$. Therefore, point w belonging to $O(a_1)$ (and nonadjacent to v) is adjacent to points $\{y,a_1,x\}$ in $O(a_1)$. Since w belongs to $O(y)$, then for the same reason as above, points w and u must be adjacent. Hence, point w is adjacent to points of $O(v)=\{u,x,a_1,y\}$. Therefore, $O(w)=O(v)$ and the subspace $A \subseteq G^2$ consisting of points $\{v,u,x,a_1,y,w\}$ is the minimal two-sphere $A=S^0(v,w) \oplus S^0(u,a_1) \oplus S^0(x,y)$. Since $A \subseteq G^2$, then $A=G^2$ by proposition 3.3 and $G^2$ is the minimal two-sphere. Therefore, $H^2$ is the two-sphere by corollary 4.2. The proof is complete. □

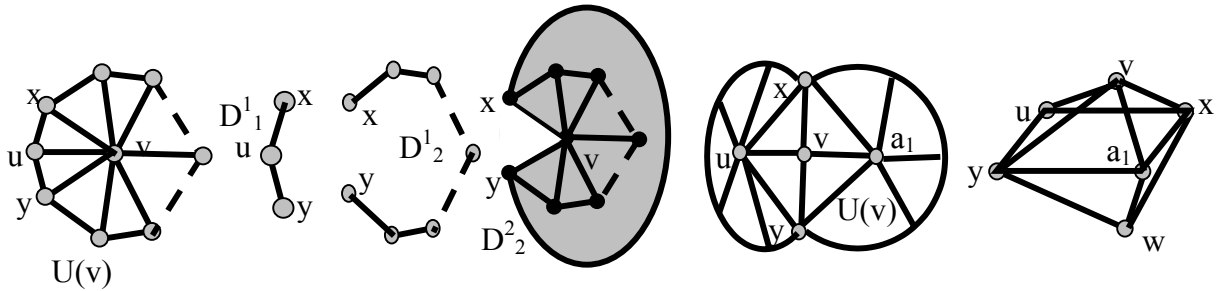

Figure 9. The rim of point v converts into the minimal one-dimensional sphere $S^1(4)=\{x,a_1,y,u\}$ and $G^2$ converts into the minimal two-dimensional sphere $S^2(6)=\{v,x,a_1,y,u,w\}$.

Both proofs are implicitly based on the assumption that $H^2$ has a finite number of points. h-Transformations reduce the number of points to a certain amount, which allows us to make conclusion about the structure of $H^2$. This is an important difference between a continuous and a digital approach.

The rim of point 4 of the two-dimensional projective plane $P^2$ depicted in fig. 5.2, is a one-sphere $\{1,3,5,7,9,6\}=D^1_1 \# D^1_2$. For $D^1_1$, there is a two-disk $D^2_1$ which contains $D^1_1$. Therefore, $D^1_1$ is embedded in $P^2$. But it can be checked directly that $D^1_2$ is not embedded.. Therefore P is not the two-sphere according to theorem 5.2.

Theorems 5.2 says that a closed two-manifold we have to check only local one-disks belonging to the rims of points.

6. Three-manifolds.

Since in this part, we are concerned with two-disks $D^2$, we introduce conditions which allow us to transform two-disks belonging to closed n-manifolds without changing topological features of n-manifolds such as the connectedness, the dimension, the Euler characteristic and the homology groups. Therefore, two-disks should be transformed by h-transformations. According to definition 4.3, it can be done if there is an n-disk $D^n$ such that $\partial D^2\subseteq \partial D^n$, $IntD^2\subseteq IntD^n$. In this case, we can replace $D^n$ by $v\oplus\partial D^n$ and $D^2$ by $v\oplus\partial D^2$ (fig. 10).

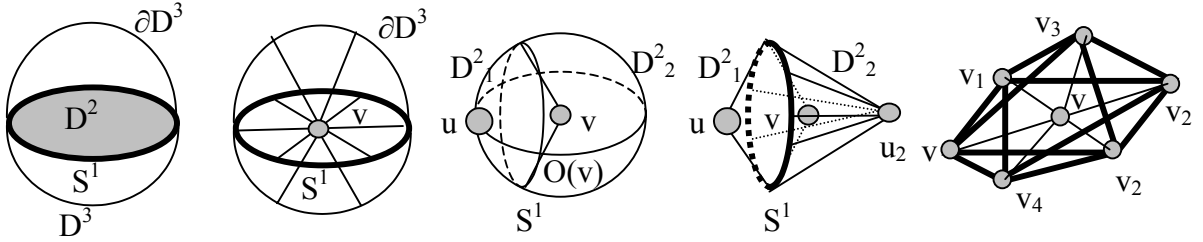

Figure 10. $S^1$ is a one-sphere, $D^2$ is a two-disk, $S^1=\partial D^2$, $D^3$ is a three-disk, $S^1\subseteq\partial D^3$ and $IntD^2\subseteq IntD^3$.

Definition 6.1.
Let $H^n$ be a closed n-manifold and $D^k$, $k<n$, be a k-disk belonging to $H^n$. $D^k$ is called embedded in $H^n$ if there is an n-disk $D^n$ such that $\partial D^k\subseteq \partial D^n$ and $IntD^k\subseteq IntD^n$.

Theorem 6.1.
Let $H^3$ be a closed three-manifold such that any two-disk $D^2\subseteq H^3$ is embedded in $H^3$. If for any one-sphere $S^1\subseteq H^3$ there is a two-disk $D^2\subseteq H^3$ such that $S^1=\partial D^2$, then $H^3$ is the three-sphere.

Proof.
Take any three-disk $D^3$ belonging to $H^3$ and replace it with the three-disk $v\oplus\partial D^3$ by the deleting of all points belonging to $IntD^3$ and the connecting of a point $v$ with any point of $\partial D^3$. This is an h-transformation that converts $D^3$ into $U(v)=v\oplus\partial D^3$. Repeat this procedure until any three-disk $D^3$ is the ball of some point $v$, $D^3=U(v)$. Denote the obtained space by $G^3$. Clearly, $G^3$ is homeomorphic to $H^3$.

For a point $v\in G^3$, take some point $u$ belonging to a two-sphere $O(v)$ (fig. 10). Then $S^1=O(v,u)=O(v)\cap O(u)$ is a one-sphere by definition 4.2, $u\oplus O(v,u)=D^2_1$ is a two-disk belonging to $O(v)$ and $O(v)-u$ also is a two-disk $D^2_2$ belonging to $O(v)$ according to lemma 4.4. Obviously, $S^1=\partial D^2_1=\partial D^2_2$ and $O(v)=D^2_1\cup D^2_2$. Since any two-disk is embedded in $G^3$ and any three-disk is the ball of a point of $G^3$ then there is a point $u_2$ such that $IntD^2_2\subseteq U(u_2)$ and $S^1=\partial D^2_2\subseteq O(u_2)$. Hence, $IntD^2_2=u_2$, $v\in O(u_2)$ and $O(v,u_2)=S^1$. Therefore, $O(v)= S^0(u,u_2)\oplus S^1$.

Take a point $v_1$ belonging to $S^1$. Using the above arguments, we find that $O(v)=S^0(u,u_2)\oplus S^0(v_1,v_2)\oplus S^0(w_1,w_2)$. Therefore, $O(v)$ is the minimal two-sphere $S^2_{min}$. Since point $v$ is chosen arbitrarily, then the rim of any point is $S^2_{min}$.

Let the rim of some point $v$ be $S^2_{min}$ consisting of points $\{v_1,v_2,\ldots v_6\}$ and consider the rim of point $v_2$ (fig. 11). $O(v_2)$ is also $S^2_{min}$ consisting of six points $\{v,v_3,v_4,v_5,v_6,w\}$. Therefore, point $w$ belonging to $O(v_2)$ (and nonadjacent to $v$) is adjacent to points $\{v_2,v_3,v_4,v_5,v_6\}$. Since $w$ belongs to $O(v_3)$, then for the same reason as above, $O(v_3)=\{v,v_1,v_2,v_5,v_6,w\}$ and point w is adjacent to points $\{v_3,v_1,v_2,v_5,v_6\}$. Hence, w is adjacent to points $\{v_1,v_2,v_3,v_4,v_5,v_6\}$. Since $O(w)$ is the minimal two-sphere, then $O(w)=O(v)=\{v_1,v_2,v_3,v_4,v_5,v_6\}$ and the subspace A consisting of points $\{w,v_1,v_2,v_3,v_4,v_5,v_6\}$ is the minimal three-sphere $A=S^0(v,w)\oplus S^0(v_1,v_2)\oplus S^0(v_3,v_4)\oplus S^0(v_5,v_6)$. Since $A\subseteq G^3$, then $G^3=A$ by proposition 3.3. Hence, $G^3$ is the minimal three-sphere and $H^3$ is the three-sphere by corollary 4.2. The proof is complete. □

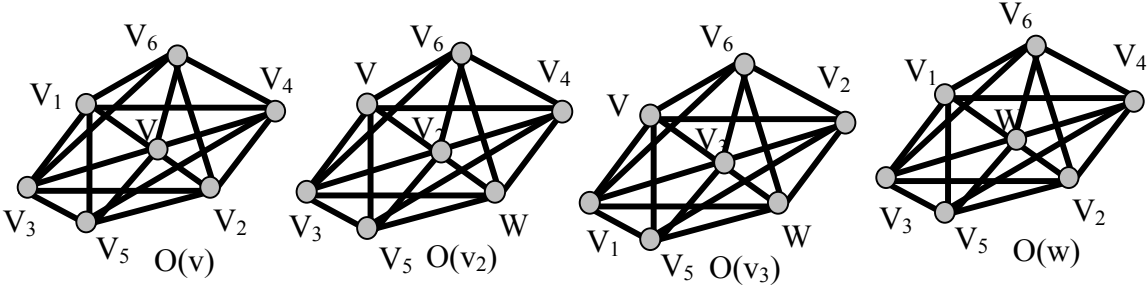

Figure 11. The rim of point w coincides with the rim of point v. Therefore, $G^3$ is a three-dimensional sphere.

Theorem 6.1 can be extended to the dimension $n>3$ in the following way.

Theorem 6.2.

Let $H^n$ be a closed n-manifold such that any (n-1)-disk is embedded in $H^n$. If for any (n-2)-sphere $S^{n-2}\subseteq H^n$ there is a (n-1)-disk $D^{n-1}\subseteq H^n$ such that $S^{n-2}=\partial D^{n-1}$, then $H^n$ is the n-sphere.

Proof.

The proof repeats the proof of theorem 6.1, if we change dimension three for dimension n. Take any n-disk $D^n$ belonging to $H^n$ and replace it with the n-disk $v\oplus\partial D^n$ by the deleting of all points belonging to $IntD^n$ and the connecting of a point v with any point of $\partial D^n$. This is an h-transformation that converts $D^n$ into $U(v)=v\oplus\partial D^n$. Repeat this procedure until any n-disk $D^n$ is the ball of some point v, $D^n=U(v)$. Denote the obtained space by $G^n$. Clearly, $G^n$ is homeomorphic to $H^n$.

For a point $v\in G^n$, take some point $u_1$ belonging to a (n-1)-sphere $O(v)$. Then $S^{n-2}=O(v,u_1)=O(v)\cap O(u_1)$ is a (n-2)-sphere by definition 4.2, $u_1\oplus O(v,u_1)=D^{n-1}_1$ is a (n-1)-disk belonging to $O(v)$ and $O(v)-u_1$ also is a (n-1)-disk $D^{n-1}_2$ belonging to $O(v)$ according to lemma 4.4. Obviously, $S^{n-2}=\partial D^{n-1}_1=\partial D^{n-1}_2$ and $O(v)=D^{n-1}_1\cup D^{n-1}_2$. Since any (n-1)-disk is embedded in $G^n$ and any n-disk is the ball of a point of $G^n$ then there is there is a point $u_2$ such that $IntD^{n-1}_2\subseteq U(u_2)$ and $S^{n-2}=\partial D^{n-1}_2\subseteq O(u_2)$. Hence, $IntD^{n-1}_2=u_2$, $v\in O(u_2)$ and $O(v,u_2)=S^{n-2}$. Therefore, $O(v)=S^0(u_1,u_2)\oplus S^{n-2}$.

Take a point $v_1$ belonging to $S^{n-2}$. Using the above arguments, we find that $O(v)=S^{n-1}=S^0(u_1,u_2)\oplus S^0(v_1,v_2)\oplus S^{n-3}$. Applying the same arguments to any point of $S^{n-3}$, we find that $O(v)$ is the join of n copies of the zero-sphere i.e., $O(v)=S^{n-1}_{min}$. Since point v is chosen arbitrarily, then the rim of any point is $S^{n-1}_{min}$.

Let the rim of some point v be $S^{n-1}_{min}$ consisting of points $\{v_1,v_2,...v_{2n}\}$ and consider the rim of point $v_2$. $O(v_2)$ is also $S^{n-1}_{min}$ and contains 2n points $\{v,v_3,v_4,...v_{2n},w\}$. Therefore, point w belonging to $O(v_2)$ (and nonadjacent to v) is adjacent to points $\{v_2,v_3,v_4,...v_{2n}\}$. Since w belongs to $O(v_3)$, then for the same reason as above, point w is adjacent to points $\{v_3,v_1,v_2,v_5,...v_{2n}\}$. Hence, w is adjacent to points $\{v_1,v_2,v_3,v_4,...v_{2n}\}$. Therefore, $O(v)\subseteq O(w)$. By proposition 3.2, $O(w)=O(v)$. Then the subspace A consisting of points $\{w,v_1,v_2,v_3,v_4,...v_{2n}\}$ is the minimal normal n-sphere $A=S^0(v,w)\oplus O(v)=S^0(v,w)\oplus S^0(v_1,v_2)\oplus S^0(v_3,v_4)\oplus...S^0(v_{2n-1},v_{2n})$. Since $A\subseteq G^n$, then $A=G^n$ by proposition 3.2. and $G^n$ is the minimal n-sphere. Therefore, $H^n$ is the n-sphere by corollary 4.2. This completes the proof. □

It is clear that on the first step in proofs of theorems in sections 5 and 6, we reduce the number of points of n-manifolds to some minimal level at least in the neighborhood of points. So, it seems desirable to study properties of such manifolds in more detail.

7. Compressed spaces.

Although in this section we deal with n-manifolds, most of the results can be applied to n-spaces in which the rim of a point is not necessarily an (n-1)-sphere.
The main difference between digital and continuous n-manifolds is that a digital n-manifold has a finite or countable number of points while a continuous n-manifold has the cardinality of the continuum. If a digital closed n-manifold has a finite number of points, it can be reduced by homeomorphic transformations while it is impossible for continuous spaces. This is essential for our further study because n-manifolds with a small number of points are easier to analyze. In the rest of the paper, we consider n-manifolds with n>0.
Definition 7.1.
A closed n-manifold N is called compressed if any n-disk D in N is the ball of some point v, D=U(v) (fig. 12).
Figure 12 shows the compressed projective plane P. The unions $U(4)\cup U(7)$ and $U(3)\cup U(5)$ are not two-disks. For the compressed two-sphere $S_1$, the union $U(1)\cup U(2)$ is not a two-disk. For the non-compressed two-sphere $S_2$, the union $U(3)\cup U(4)$ is a two-disk.
Lemma 7.1.
Any closed n-manifold N can be compressed by h-transformations.
Proof.
Note first, that the ball U(v) of any point v of N is an n-disk. Take an n-disk D belonging to N and different from the ball of any point of N. Therefore, D contains more than one interior point. Introduce connections between a point v and all point belonging to ∂D and delete all points belonging to IntD. Then N moves to a closed n-manifold M homeomorphic to N. Repeat this procedure until any n-disk is the ball of some point. If N contains the finite number of points the number of replacings is finite. This completes the proof. □
Lemma 7.2.
If N is the compressed closed n-manifold, then:
( a ) For any set of points $v_1, v_2, \ldots v_k$, the union $U(v_1)\cup U(v_2)\cup \ldots \cup U(v_k)$ of their balls is not an n-disk.
( b ) For any two adjacent points u and v, there is a one-sphere S(4) consisting of four points and containing points u and v.

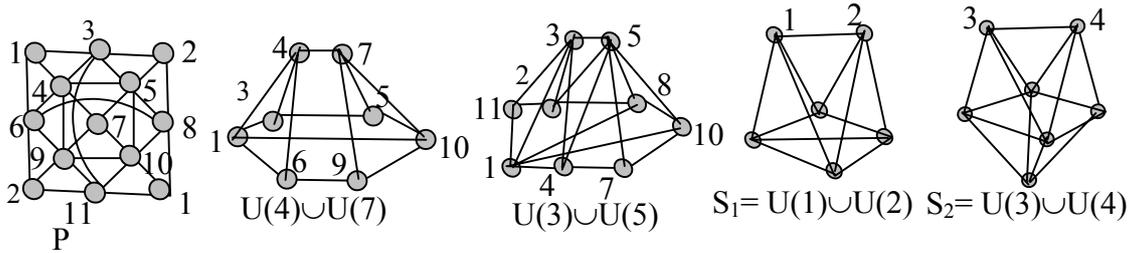

Figure 12. In the compressed two-dimensional projective plane P the unions of balls $U(4)\cup U(7)$ and $U(3)\cup U(5)$ are not two-disks. In the compressed (minimal) two-sphere $S_1$, the union of balls $U(1)\cup U(2)$ is not a two-disk. In the non-compressed two-sphere $S_2$, the union of balls $U(3)\cup U(4)$ is not a two-disk.

( c ) For any two non-adjacent points u and v, their joint rim O(u)∩O(v) is not an (n-1)-disk.
Proof.
Assertion (a) follows from definition 7.1.
Assertion (b) follows from lemma 4.10.
To prove (c), suppose that u and v are non-adjacent points such that the intersection O(u)∩O(v) of their rims is an (n-1)-disk. Then the subspace U(v)+u containing point u and all points belonging to U(v) is an n-disk. From this contradiction we conclude that O(u)∩O(v) is not an (n-1)-disk. □

Lemma 7.3.
The compressed n-sphere $S^n$ is the minimal n-sphere $S^n_{min}$.
It follows from the proof of lemma 4.3.

Lemma 7.4.
Let N be the compressed closed n-manifold. If there are adjacent points v and u such that their rims O(v) and O(u) are minimal (n-1) spheres, then N is the minimal n-sphere.
Proof.
Suppose points v and u are adjacent and O(v) and O(u) are minimal (n-1)-spheres $S^{n-1}_1$ and $S^{n-1}_2$. Then O(v,u) is the minimal (n-2)-sphere $S^{n-2}$. Therefore, O(v)=$S^0(a,u) \oplus S^{n-2}$ and O(u)=$S^0(b,v) \oplus S^{n-2}$. Since U(v)∪U(u) is not an n-disk, then by lemma 4.10, there is a one-sphere $S^1(4)$ containing points v and u and consisting of four points. Therefore, $S^1(4)$ is necessarily {v,u,a,b}, where points a and b are adjacent. Hence,
A=U(v)∪U(u)=$S^)(a,u) \oplus O(u)=S^)(a,u) \oplus S^0(b,v) \oplus S^{n-2}$ is the minimal n-sphere. Since A⊆N, then A=N by proposition 3.3. Hence, N is the minimal n-sphere. This completes the proof. □

Corollary 7.1.
Let N be the compressed closed n-manifold.
( a ) If there is a point v∈N such that O(v) is not the minimal (n-1)-sphere, then N is not the n-sphere.
( b ) If there is a one-sphere $S^1 \subseteq N$ such that there is no point v adjacent to all points of $S^1$, then N is not the n-sphere.

8. Concluding remarks and summary of results.

Possibly, there is a parallel between the Poincaré conjecture for digital and continuous three-manifolds. According to J. Milnor [15], the Poincaré conjecture can be formulated as follows:
If a smooth compact 3-dimensional manifold M has the property that every simple closed curve within the manifold can be deformed continuously to a point, does
it follow that M is homeomorphic to the three-sphere S ?
Or in another formulation:
If a closed three-dimensional manifold has trivial fundamental group, must it be homeomorphic to the three-sphere?
Intuitively speaking, if a closed curve can be contracted continuously to a point, then this curve is the boundary of some two-disk. It seems natural for a continuous closed three-manifold M that for a two-dimensional disk $D^2$ belonging to M there is a three-dimensional disk $D^3$ such that $\partial D^2 \subseteq \partial D^3$ and $Int D^2 \subseteq Int D^3$. We use this requirement in the digital form in theorem 6.1.
Speaking about the link between digital and continuous spaces, there is a natural way of obtaining the digital model of a continuous space. That is to take a finite or countable cover F of a space N by disks and to build the intersection graph G(F) of this cover [9].

G(F) is considered as the digital model of N. It was proven that if F={F$_1$,F$_2$,…} and H={H$_1$,H$_2$,…} are complete covers of N, then the intersection graphs G(F) and G(H) are, at least, homotopic. If N is a continuous n-dimensional manifold and G(F) and G(H) are digital n-manifolds, then G(F) and G(H) are homeomorphic because homeomorphic transformations are contractible transformations retaining the dimension. Using results of parts 5 and 6, we can determine whether a digital n-manifold M is the digital n-sphere. For example, let N be a continuous three-dimensional manifold, F={F$_1$,F$_2$,…} be the complete cover of N and G(F) be the intersection graph of this cover. Suppose that G(F) can be converted to the digital minimal three-sphere. Then for any complete cover H={H$_1$,H$_2$,…}, G(H) can be contracted only to the digital minimal three-sphere and there is the complete cover of N consisting of eight continuous thee-dimensional disks. On this basis, we can conclude that N is a continuous three-dimensional sphere.

Classification of three-manifolds.

We end with the problem for further study: In this paper as an example, we consider the two-sphere S and the two-dimensional projective plane P and show directly that S-v is homotopic to a point and P-v is homotopic to a one-sphere. If for any closed digital n-manifold M, M-v can be converted to a digital space (not necessarily normal) whose dimension is strictly smaller than n it can be used for classification of digital and continuous n-manifolds on the basis of classification of digital (n-1)-spaces. Another advantage of digital approach is that digital spaces have a finite or countable number of points, which also can help classification problem. On first step, we can classify three manifolds with a finite number of points by the amount of points that their compressed models have. For example, the compressed three sphere has eight points, the minimal two-sphere has six points, the minimal two-dimensional projective plane has eleven points and the minimal two-dimensional torus has sixteen points.

Here are some of the results established in this paper.
- Let $H^n$ be a closed n-manifold. If there is a point v such that $H^n$-v is an n-disk, then $H^n$ is the n-sphere.
- Let $H^2$ be a closed two-manifold. If for any one-sphere $S^1$ belonging to $H^2$ there is a two-disk $D^2$ belonging to $H^2$ such that $S^1=\partial D^2$, then $H^2$ is the two-sphere.
- Let $H^3$ be a closed three-manifold. If for any one-sphere $S^1 \subseteq H^3$ there is a two-disk $D^2 \subseteq H^3$ such that $S^1=\partial D^2$ and any two-disk $D^2 \subseteq H^3$ is embedded, then $H^3$ is the three-sphere.

References.